\documentclass[reprint,twocolumn,floats, floatfix,superscriptaddress,amsmath,amssymb,aps]{revtex4-1}

\usepackage{graphicx}
\usepackage{dcolumn}
\usepackage{bm}
\usepackage{color}
\usepackage{lipsum}

\begin{document}

\title{Wrinkling instability in unsupported epithelial sheets}

\author{Ur\v ska Andren\v sek}
\affiliation{Faculty of Mathematics and Physics, University of Ljubljana, Jadranska 19, SI-1000 Ljubljana, Slovenia}
\affiliation{Jo\v zef Stefan Institute, Jamova 39, SI-1000 Ljubljana, Slovenia}%

\author{Primo\v z Ziherl}
\affiliation{Faculty of Mathematics and Physics, University of Ljubljana, Jadranska 19, SI-1000 Ljubljana, Slovenia}
\affiliation{Jo\v zef Stefan Institute, Jamova 39, SI-1000 Ljubljana, Slovenia}%

\author{Matej Krajnc}
\email{matej.krajnc@ijs.si}
\affiliation{Jo\v zef Stefan Institute, Jamova 39, SI-1000 Ljubljana, Slovenia}%

\begin{abstract}
We investigate the elasticity of an unsupported epithelial monolayer and we discover that unlike a thin solid plate, which wrinkles if geometrically incompatible with the underlying substrate, the epithelium may do so even in absence of the substrate. From a cell-based model, we derive an exact elasticity theory and discover wrinkling driven by the differential apico-basal surface tension. Our theory is mapped onto that for supported plates by introducing a phantom substrate whose stiffness is finite beyond a critical differential tension. This suggests a new mechanism for an autonomous control of tissues over the length scale of their surface patterns.
\end{abstract}

\maketitle

{\it Introduction.---}Complex shapes and patterns often arise even in simple physical setups~\cite{cerda03,klein07,efrati09}. For instance, a uniaxial in-plane compression makes a thin plate attached to an elastic foundation wrinkle due to the competition between the bending energy of the plate and the bulk elastic energy of the substrate~\cite{pocivavsek08,brau11}. This competition yields the 
length scale of wrinkles, which scales with the ratio of the stiffness coefficient of the foundation $K$ and the bending modulus of the plate $B$ as~\cite{brau11,oshri15,brau13}
\begin{equation}
	\label{eq:1}
	\lambda_0\sim\left (\frac{K}{B}\right )^{-1/3}\>.
\end{equation}
In proliferating epithelia, cell division causes areal growth of the tissue relative to the underlying stroma. The proliferation-driven differential growth leads to a geometric incompatibility between the epi\-thelium and the stroma, triggering wrinkling with the wavelength given by Eq.~(\ref{eq:1}) much like an external compressive force~\cite{hannezo11,shyer13}.

Periodic surface patterns in epithelia may also arise from the intrinsic curvature due to apico-basal differential surface tension, which is known to drive characteristic local deformations such as epithelial curling and buckling {\it in vitro}~\cite{Luu11,wyatt19,fouchard20} as well as folding during morphogenesis and organogenesis~\cite{sui18,garcia19,rozman20}. This mechanism may be particularly relevant in unsupported epithelia, whose deformations mostly rely on internal active stresses. We previously investigated such cases and we showed 
that the differential tension alone can produce periodic deformation patterns even if the tissue is not attached to a substrate~\cite{krajnc13,krajnc15,storgel16}. However, these stu\-dies only considered the large-deformation regime where the equilibrium wavelength is mainly determined by steric interactions between cells, and they neither identified the parameters that control the 
wavelength of shape patterns at the onset of wrinkling nor did they show why epithelia wrinkle in absence of the substrate in the first place.

\begin{figure}[b!]
\vspace*{-4mm}
    \includegraphics[]{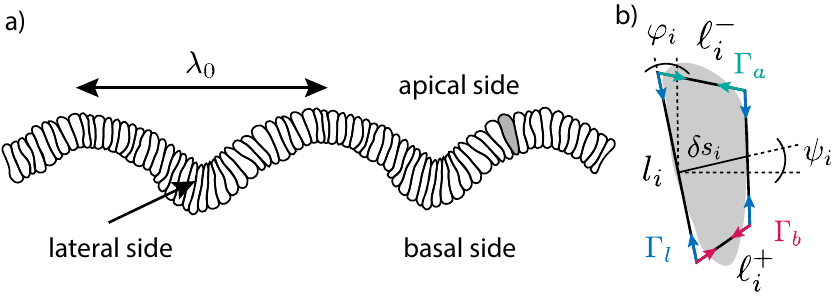}
\vspace*{-3mm}
    \caption{(a)~Waveform of a wrinkled epithelial monolayer, its wavelength denoted by $\lambda_0$. (b)~Schematic of a cell cross section, parametrized by two lengths ($l_i$ and $\delta s_i$) and two angles ($\varphi_i$ and $\psi_i$). The line tensions of the basal, apical, and lateral sides read $\Gamma_b$, $\Gamma_a$, and $\Gamma_l$, respectively. }
    \label{F1}
\end{figure}

Here we address these issues by deriving an elasti\-city theory from the cell-level mechanics based on surface tension. Our theory describes a new type of tension-controlled transition whereby tissues switch between the buckling instability, where they buckle out of plane, and the wrinkling instability, where their deformation pattern is characterized by a finite wavelength. We show that the effect of surface tensions can be translated into the stiffness of a phantom bulk substrate, which becomes finite at the critical point. This concept reconciles the wrinkling of unsupported epithelia with
that of thin solid plates supported by substrates, and it serves as a basis of a novel mechanism for autonomous control of shape patterns in epithelia.

{\it The model.---}Our model tissue is represented by a chain of quadrilateral cell cross sections~(Fig.~\ref{F1}a). The basal, apical, and lateral cell sides are under effective line tensions $\Gamma_b$, $\Gamma_a$, and $\Gamma_l$, respectively, which account for cortical tension, apico-basal differential surface tension, describing apico-basal tension polarity, and cell-cell adhesion. Following previous models, we view all three terms as strain-independent and we model them by constant line tensions~\cite{derganc09,krajnc13,storgel16}. The cells are assumed incompressible, and therefore their cross-section areas are fixed and equal in all cells ($A_i=A_0$ for all $i$). 

We parametrize the shape of $i$th cell by the  lengths of its lateral sides $l_i$ and $l_{i+1}$, cell width $\delta s_i$, the orientation of lateral sides relative to the vertical axis given by angles $\varphi_i$ and $\varphi_{i+1}$, respectively, and the orientation of cell midline relative to the horizontal axis given by angle $\psi_i$~(Fig.~\ref{F1}b). The constant-cell-area constraint allows us to express $\delta s_i$ in terms of all other variables, so that in dimensionless form, where energy and lengths are given
in units of $\Gamma_l\sqrt{A_0}$ and $\sqrt{A_0}$, respectively, the energy of a single cell reads
\begin{equation}
    \label{vertMod}
    w_i=\frac{\Gamma_b}{\Gamma_l}\ell_i^{+}+\frac{\Gamma_a}{\Gamma_l}\ell_i^{-}+\frac{1}{2}\left (l_i+l_{i+1}\right )\>,
\end{equation}
where the lengths of basal and apical cell sides $\ell_i^+$ and $\ell_i^-$, respectively, depend on $l_i$, $l_{i+1}$, $\varphi_i$, $\varphi_{i+1}$, and $\psi_i$~(Supplemental Material, Sec.~I~\cite{suppl}). Next we define the average and the differential apico-basal surface tension
\begin{equation}
    \Gamma=\frac{\Gamma_a+\Gamma_b}{\Gamma_l}\>\>{\rm and}\>\>\Delta=\frac{\Gamma_a-\Gamma_b}{\Gamma_l}\>,
\end{equation}
respectively, and we study homogeneous tissues where $\Gamma$ is the same in all cells and so is $\Delta$. The energy of the tissue is a sum of single-cell terms: $W=\sum_i^{N} w_i$.

{\it Epithelial elasticity.---}Unlike the classical plate theory which stems from the strain distribution within the plate, the elasticity of our model tissue arises from surface mechanics. Interestingly, this affects the scaling of the 
stretching and bending moduli ($K_s$ and $B$, respectively) with tissue thickness $l_0$. Considering only pure stretching and bending deformation modes, we find that $K_s\sim l_0^2$ and $B\sim 1$, which 
differs from the well-known results for solid plates where $K_s^{\rm plate}\sim l_0$ and $B^{\rm plate}\sim l_0^3$~(Supplemental Material, Sec.~II~\cite{suppl}). 
Since wrinkles result from the competition between the two deformation modes, this scaling suggests that epithelial wrinkling is phenomenologically distinct from that in plates. 

To derive the exact elastic theory, we consider a deformation from the flat reference state where $l_i=l_0=\sigma_0^{-1}$, $\varphi_i=0$, and $\psi_i=0$; here $\sigma_0=\Gamma^{-1/2}$ is the reference cell width.
We view $l_i$, $\delta s_i$, $\varphi_i$, and $\psi_i$ as continuous functions of the distance $\sigma$ along the refe\-rence-state midline so that $l_i\to l(\sigma)$ etc.,  and we relate their values in the $(i+1)$th and $i$th cell by their derivatives with respect to $\sigma$: 
$l_{i+1}\approx l(\sigma)+\dot l(\sigma)\sigma_0$ etc. In the continuum limit, the sum over cells is replaced by the integral over the reference-state midline
and the energy reads 
$W=\int_0^{N\sigma_0}\mathcal L{\rm d}\sigma=\Gamma^{1/2}\int_0^{N\sigma_0}\left [w\left( l,\varphi,\psi,\dot l,\dot\varphi\right )+\mu\delta s\left( l,\varphi,\psi,\dot l,\dot\varphi\right )\cos\psi\right ]{\rm d}\sigma$. Here, the Lagrange multiplier $\mu$ represents the compressive force. 

We expand the Lagrangian density $\mathcal L$ for small deviations from the reference state up to the second order and we obtain 
\begin{equation}
    \begin{split}
        \mathcal L\approx &2\Gamma+\mu-\frac{\mu}{\sqrt{\Gamma} }\delta l - \frac{\Delta \sqrt{\Gamma}}{2}\dot\varphi +\frac{\Gamma-\mu}{2} \left(\psi-\varphi \right)^2+\\ 
        &+\frac{\mu}{2}\psi ^2 +\frac{\Gamma+\mu}{\Gamma}\delta l^2+\delta l \left(\frac{1}{\sqrt{\Gamma}}\delta\dot l - \frac{\Delta}{2} \dot\varphi\right)+\\
        &+\left(\psi-\varphi \right) \left(\frac{\Delta}{2} \delta\dot l - \frac{\sqrt{\Gamma}}{2} \dot\varphi \right)-\frac{ \mu}{2 \sqrt{\Gamma} }\psi \dot{\varphi}-\\
        &-\frac{\Delta}{2\sqrt{\Gamma}} \delta\dot l\dot\varphi+\frac{2(\mu+\Gamma) + \Gamma^3}{8 \Gamma ^{2}} \delta\dot l^2     +\frac{\mu+\Gamma}{4\Gamma}\dot\varphi^2\>,
    \end{split}
\end{equation}
where $\delta l=l-l_0$ is the deviation of the late\-ral-side length from the reference value. Next we spell out the Euler--Lagrange equations~(Supplemental Material, Sec.~III~\cite{suppl}) and Fourier-transform $\delta l(\sigma), \varphi(\sigma)$, and $\psi(\sigma$): $\delta l(\sigma)=\int \delta \widetilde l(q)\exp{({\rm i}q\sigma)}{\rm d}q/2\pi$,
etc.
where $q=2\pi/\lambda$. These transformations reduce the Euler--Lagrange equations to a single algebraic equation for the wavenumber of the instability:
\begin{equation}
	\label{eq:q}
    q_0^4+\widetilde Fq_0^2+\widetilde K=0\>,
\end{equation}
where $\widetilde F$ and $\widetilde K$ are functions of $\Gamma$, $\Delta$, and $\mu$~(Supplemental Material, Sec.~IV~\cite{suppl}). Surprisingly, even though our epi\-thelial sheets are neither thin (as their thickness is gene\-rally not negligible compared to the wrinkle wavelength) nor supported by a substrate, Eq.~(\ref{eq:q}) is equi\-valent to the version of Eq.~(7) in Ref.~\cite{brau13} which describes wrinkling of thin plates on a liquid foundation~\cite{footnote1}. By this analogy, $\widetilde F$ and $\widetilde K$ can be viewed as an effective in-plane force and an effective stiffness of a {\it phantom} substrate, respectively, both expressed relative to the bending mo\-dulus of the epithelium.

\begin{figure}[h!]
    \includegraphics[]{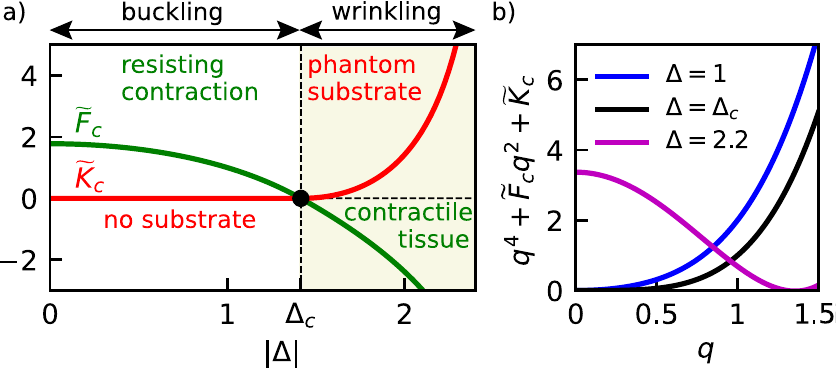}
    \caption{(a)~$\widetilde{F}_c(\Delta)$ (green curve) and $\widetilde{K}_c(\Delta)$ (red curve). (b) Left-hand side of Eq.~(5) for $\Delta=1$, $\Delta_c(=\sqrt{2})$, and~$2.2$ (blue, black, and purple curves, respectively). In both panels,~$\Gamma=4$. }
    \label{F2}
\end{figure}

Our system becomes unstable at a critical in-plane compressive force $\mu=\mu_c$~(Supplemental Material, Sec.~IV~\cite{suppl}), with the mode of instability switching from buckling to wrinkling as the magnitude of the differential surface tension $|\Delta|$ is increased beyond $\Delta_c=\sqrt{2}$. This transition is nicely reflected in the critical values of the two effective parameters $\widetilde F_c$ and $\widetilde K_c$ corresponding to $\mu_c$. In particular, for $\left |\Delta\right |<\Delta_c$ the critical in-plane force $\widetilde F_c>0$ whereas the critical substrate stiffness $\widetilde K_c=0$ so that the phantom substrate is absent~(Fig.~\ref{F2}a). In this regime, Eq.~(\ref{eq:q}) simplifies to $q_0^2\left (q_0^2+\widetilde F_c\right )=0$ and its only real solution $q_0=0$ implies buckling under compression~(Fig.~\ref{F2}b) as expected for unsupported plates.

\begin{figure*}[htb!]
    \includegraphics[width=18.cm]{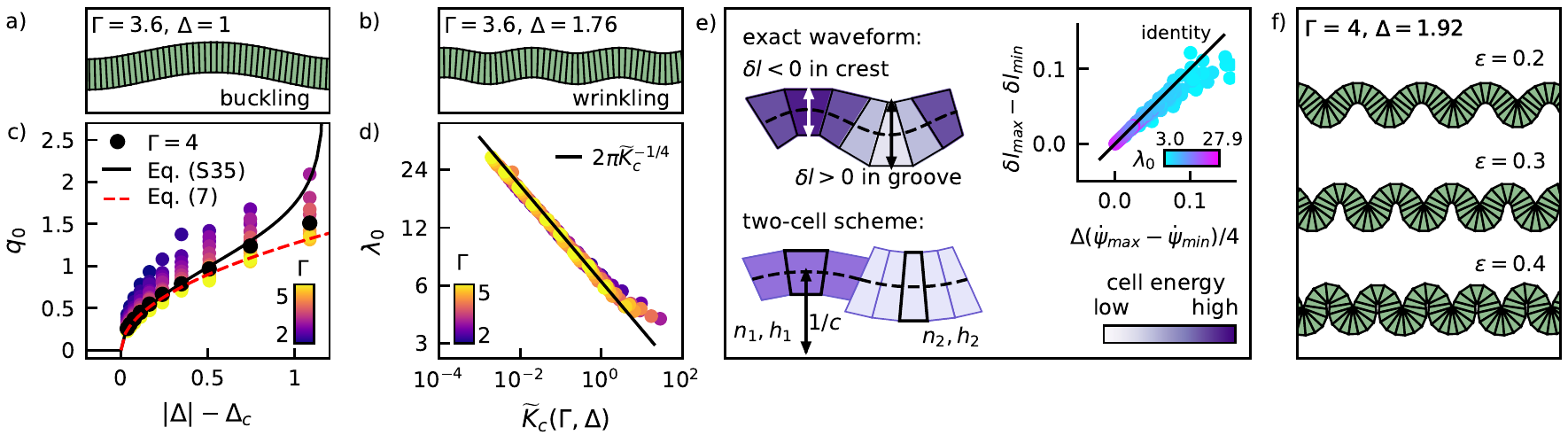}
    \caption{(a,~b)~Buckled and wrinkled shapes at $\Delta=1$~(a) and $\Delta=1.76$~(b); $\Gamma=3.6$ in both cases. (c)~Numerically obtained equilibrium wavenumber $q_0$ vs.~$\left |\Delta\right |-\Delta_c$. Solid and dashed lines show the exact result~[Eq.~(S35)] and the result for small $\left |\Delta\right |-\Delta_c$~[Eq.~(\ref{eq:q0})], respectively. 
    (d)~Equilibrium wavelength $\lambda_0$ vs.~critical phantom-substrate stiffness~$\widetilde K_c$ given by Eq.~(S33). (e)~Cell-height modulation in a tissue with $\Delta>\Delta_c$~(top) and two-cell scheme~(bottom). Inset: Cell-height modulation vs.~curvature modulation; solid line is identity predicted by Ref.~\cite{footnote2}. Datapoints deviate from the identity at small $\lambda_0$, where the discrete nature of tissue becomes more pronounced. (f)~Compressed tissues for large strains $\epsilon=0.2,\>0.3,$ and $0.4$.}
    \label{F3}
\end{figure*}

However, at $\left |\Delta\right |>\Delta_c$, $\widetilde F_c$ is negative (indicating a contractile tissue) whereas $\widetilde K_c$ is positive, which signals the presence of a phantom substrate~(Fig.~\ref{F2}a). In this regime, Eq.~(\ref{eq:q}) has a double real solution at a critical in-plane compressive force $\mu_c$ (Fig.~\ref{F2}b), implying $\widetilde F_c=-2\widetilde K_c^{1/2}$. Thus, Eq.~(\ref{eq:q}) reduces to $\left (q_0^2-\widetilde K_c^{1/2}\right )^2=0$ and its solution $q_0=\widetilde K_c^{1/4}$ describes wrinkles with a wavelength
\begin{equation}
	\label{eq:l0}
	\lambda_0=2\pi\widetilde K_c^{-1/4}\>.
\end{equation}
This result agrees with that for thin elastic plates supported by a thick {\it liquid} foundation~\cite{brau13}. Importantly, the scaling exponent in Eq.~(\ref{eq:l0}) differs from $-1/3$ obtained in the case where the epithelium is treated as a thin plate attached to a thick elastic substrate~(Eq.~(\ref{eq:1}) and Ref.~\cite{hannezo11}). 

Interestingly, the critical external in-plane force $\mu_c$ also changes sign at $\left |\Delta\right|=\sqrt{2}$, such that an extensile force is required in the wrinkling regime so as to prevent tissue collapse. If a collapse {\it is} allowed, the equilibrium wavelength is determined by steric repulsion and is of the order of twice the cell height~($\lambda_0\sim2\sigma_0^{-1}$) and almost independent of~$\Delta$~\cite{krajnc13}.

{\it Vertex model.---}To address cell-level mechanisms of the buckling-to-wrinkling transition, we next employ the vertex model where the tissue shape is parametrized by the positions of cell vertices. The equilibrium wrinkled and buckled states are computed by minimizing the total ener\-gy $W=\sum_i^{N}w_i$ and the size of the simulation box is varied so as to find the critical compressive force at which the instability occurs~(Figs.~\ref{F3}a, b and Supplemental Material, Sec.~V~\cite{suppl}).

We compute the equilibrium wavelength $\lambda_0$ for a wide range of parameters $\Gamma$ and $\Delta$. By interpreting $q_0=2\pi/\lambda_0$ and $\left |\Delta\right |$ as an order and a control parameter, respectively, we find a second-order buckling-to-wrinkling transition at $\left |\Delta\right |=\Delta_c=\sqrt{2}$. We compare the results to the elastic theory by plotting the wavenumber of wrinkles $q_0$ versus the control parameter $|\Delta|$ and in the regime where the wavelengths are much longer than the typical cell size and the continuum approximation is justified, we find perfect agreement~(Fig.~\ref{F3}c). The data also agree with the critical scaling of the wavenumber
\begin{equation}
    \label{eq:q0}
    q_0\approx\frac{2^{5/4}}{\sqrt{\Gamma-2/\Gamma}}\left (\left|\Delta\right |-\Delta_c\right )^{1/2}\>,
\end{equation}
obtained by studying our theory around $|\Delta |=\sqrt{2}$. Furthermore, the data can be collapsed onto a universal curve given by Eq.~(\ref{eq:l0}) by plotting the equilibrium wavelength versus the critical substrate stiffness $\widetilde K_c$ given by Eq.~(S34)~(Fig.~\ref{F3}d). This confirms that in the context of wrinkling, the surface tensions of the cells give rise to mechanics that are indistinguishable from the bulk elasticity of thick substrates.

{\it Cell-height modulation.---}The origin of the wrinkling instability can be intuitively understood by inspecting the shape of individual cells along the waveform. This shows that cell height is modulated in agreement with the elastic theory, which predicts that the modulation is proportional to the local curvature $\dot\psi(\sigma)$: $\delta l(\sigma)\approx\Delta\dot\psi(\sigma)/4$~(Fig.~\ref{F3}e and Ref.~\cite{footnote2}). Since cells are incompressible, this modulation implies symmetry breaking between the number density of cells in groove (where $\dot\psi>0$) and that in the crest (where $\dot\psi<0$), such that the ener\-getically favorable cells in the segment with the preferred local curvature are packed more densely than elsewhere. For instance, for $\Delta>0$ the low-energy cells in the groove are taller and narrower than the high-energy cells in the crest~(Fig.~\ref{F3}e). We note that in our model, the phase of cell-height modulation depends on the sign of $\Delta$ and is not universal like in a wrinkled supported solid plate where the grooves are always thinner than the crests~\cite{Holland18} and in a fluid-like films anchored to a solid substrate by a fibrous scaffold where it is the opposite~\cite{Engstrom18}.

In our model, cell-height modulation is an emergent phenomenon that occurs even though the properties of the cells are homogeneous across the epithelium. This implies that the wrinkling instability cannot be predicted from the behavior of individual cells whose preferred shape is a trapezoid with height $h_0=\Gamma^{1/2}$ and curvature $c_0\approx -2\Gamma^{1/2}\Delta$~\cite{storgel16}. In line with the Gauss-Bonnet theorem, a sheet described by an in-plane periodic curve which carries a bending energy ${\rm d}w_b/{\rm d}s=k(c-c_0)^2/2$ favors a flat configuration despite the spontaneous curvature $c_0$.

To explain the loss of stability in a flat tissue as simply as possible, we consider a waveform consisting of two circular arcs of curvatures $-c$ and $+c$ (i.e., groove and the crest, respectively) subtending an angle~$\phi$~(Fig.~\ref{F3}e). All groove cells are identical and so are those in the crest; we denote the two cell types by indices 1 and 2. 
The total energy of the waveform, is a weighted sum of single-cell energies $w_1$ and $w_2$: $W=N(h_1w_1+h_2w_2)/(h_1+h_2)$~(Supplemental Material, Sec.~VI~\cite{suppl}). In agreement with our elasticity theory, the energy is minimized when groove and crest cells assume different heights ($h_1\neq h_2$), thereby breaking the groove-crest symmetry. The height difference agrees with the magnitude of cell-height modulation from the elasticity theory: $h_1-h_2=c\Delta/2$. Furthermore, this minimal scheme gives the exact critical point $\left |\Delta\right |=\sqrt{2}$, where the difference of the energies of the model waveform and the flat tissue $\delta W=N(2-\Delta^2)c^2/(16\Gamma^{1/2})$ changes sign. In all, this claculation demonstrates that the emergent groove-crest asymmetry in cell height is the dominant mechanism of the wrinkling instability. Previous theories of epi\-thelial elasticity captured this effect to a certain extent, but they did not manage to reproduce the instabi\-lity due to geometric oversimplifications~\cite{krajnc15,haas19}.

To check whether the exact mapping of epithelial elasticity to the plate-substrate system~[Eq.~(\ref{eq:q})] also holds for large deformations, we investigate tissue shapes at large strains; we note that the vertex model includes the nonlinear mechanics absent in our harmonic theory. Unlike thin plates which undergo period bifurcations and wrinkle-to-fold transition when supported by solid and liquid substrates, respectively~\cite{pocivavsek08,brau11,brau13}, we find that epi\-thelia remain wrinkled with a single mode even at large deformations~(Fig.~\ref{F3}f). The absence of the wrinkle-to-fold transition, which appears in thin plates attached to fluid substrates, can be intuitively understood from the point of view of cell-height modulation. The groove-crest  asymmetry allows the tissue side subjected to a higher surface tension to decrease its area compared to the opposite, less tense, side. Compared to the wrinkled configuration, this effect would be much less pronounced in a fold configuration, where most of the tissue would be flat with cells having apical and basal sides of equal size.

%
%

\begin{figure}[t!]
    \includegraphics[]{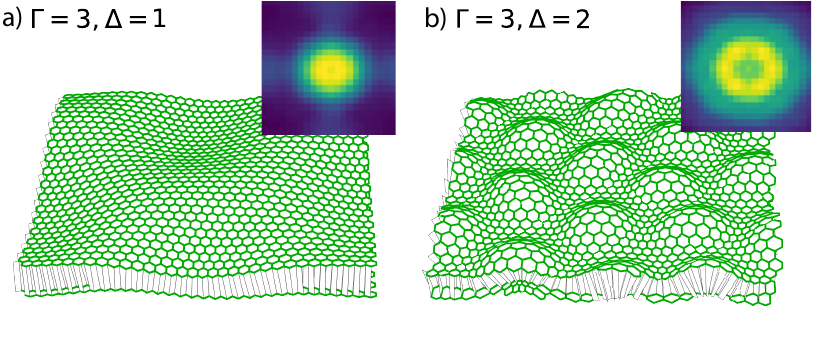}
    \caption{Epithelial sheets at $\Gamma=3$ for $\Delta=1$~(a) and $\Delta=2$~(b) at isotropic in-plane strain of $\epsilon=0.02$. Insets show Fourier transforms of the apical surfaces, averaged over 100 instances obtained from randomly perturbed flat initial configurations.}
    \label{F4}
\end{figure}

{\it 3D model.---}Our reduced-dimensionality model consi\-ders the cross section of the tissue along the waveform, effectively assuming a fixed cell dimension in the perpendicular direction. To show that the buckling-to-wrinkling transition is not an artifact of this assumption, we 
employ a 3D vertex model~\cite{rozman21} where the epithelium is represented by a sheet-like packing of six-coordinated polyhedral cells---and we observe buckled states at small $\Delta$ and wrinkled states at large $\Delta$ in agreement with the 2D model. This is illustrated in Figs.~\ref{F4}a and b which show that depending on $\Delta$, a small isotropic in-plane compression results either in a buckled state characterized by a wavenumber defined by patch size $D$ (Fig.~\ref{F4}a) or in a wrinkled state with a well-defined $q$ larger than $2\pi/D$ (Fig.~\ref{F4}b). A detailed analysis of the transition including the eva\-luation of the critical differential surface tension carried out in a model that permits topological changes is beyond the scope of this work and is relegated to a forthcoming publication.

{\it Discussion.---}Our elasticity theory of unsupported epithelial monolayers demonstrates that intraepithelial tensions drive the formation of wrinkles similar to those observed in thin plates supported by thick substrates even though the underlying physics relies solely on surface mechanics and includes no bulk deformations. These results are important because they show that a sheet-like biological tissue can wrinkle even when its mechanical interaction with the environment is weak, say because the adjacent substances are gel-like and cannot sustain long-term shear stresses. Our model suggests that these effectively unsupported epithelial tissues, which appear, e.g., in some early embryos~\cite{Wolpert19} and in va\-rious tissue-derived epithelial organoids~\cite{Sato09,Zietek15,Jung11,Huch13,Barker10,Greggio13}, can autonomously control their surface patterns just as if they rested on a substrate.


At the cell scale, epithelial wrinkling is based on the breaking of the groove-crest symmetry due to cell-height modulation along the waveform~(Fig.~\ref{F3}e), which was 
observed in 
cell monola\-yers cultured on wavy substrates~\cite{luciano21,harmand22} and described using the same microscopic cell-level framework as employed here~[Eq.~(\ref{vertMod})].

According to our results, the wrinkling wavelength can range between less than $\sim10$ and $\sim100$ cell sizes when the apico-basal differential tension $\left |\Gamma_a-\Gamma_b\right |$ is comparable to the lateral tension $\Gamma_l$~(Fig.~\ref{F3}c). Given that the surface tensions are often of similar magnitude as shown, e.g., in Ref.~\cite{sui18}, our proposed wrinkling mechanism
can be readily studied 
either {\it in vitro} or {\it in vivo}---say using optogenetic tools or genetic mani\-pulations~\cite{martinezara22} and possibly employing explants~\cite{Luu11}---so as to verify its postulated role in embryonic and organoid morphogenesis. If confirmed, our mechanism would complement existing theories of tissue wrinkling including constrained expansion~\cite{hannezo11,brau13,shyer13,Tallinen16,Balbi20}, packing of cell nuclei~\cite{Karzbrun18}, and the buckling-without-bending effect~\cite{Engstrom18}.


We thank Eric Wieschaus, Jan Rozman, Simon Godec, Miha Brojan, Jan Zavodnik, Andrej Ko\v smrlj, Edouard Hannezo, and the members of the Theoretical Biophysics Group at Jo\v zef Stefan Institute for fruitful discussions. We acknowledge the financial support from the Slovenian Research Agency (research project No.~J1-3009 and research core funding No.~P1-0055).

\begin{thebibliography}{99}
\bibitem{footnote1}
    Equation~(7) in Ref.~\cite{brau13}, $By^{(4)}+F\ddot y+Ky=0$, is a ge\-neral force-balance equation for a thin plate resting on a bulk substrate. By inserting the ansatz $y(s)=y_0\sin{(qs)}$, one obtains an algebraic equation for the wavenumber, which reads $Bq^4+Fq^2+K=0$, where $K$ is the substrate's stiffness. In case of a liquid substrate, $K=\rho g$ is an effective stiffness due to fluid's weight; here $\rho$ and $g$ are the fluid's mass density and gravitational acceleration, respectively.
\end{thebibliography}

\end{document}



\title{Wrinkling instability in unsupported epithelial sheets:\\Supplemental material}

\author{Ur\v ska Andren\v sek}
\affiliation{Faculty of Mathematics and Physics, University of Ljubljana, Jadranska 19, SI-1000 Ljubljana, Slovenia}
\affiliation{Jo\v zef Stefan Institute, Jamova 39, SI-1000 Ljubljana, Slovenia}%

\author{Primo\v z Ziherl}
\affiliation{Faculty of Mathematics and Physics, University of Ljubljana, Jadranska 19, SI-1000 Ljubljana, Slovenia}
\affiliation{Jo\v zef Stefan Institute, Jamova 39, SI-1000 Ljubljana, Slovenia}%

\author{Matej Krajnc}
\email{matej.krajnc@ijs.si}
\affiliation{Jo\v zef Stefan Institute, Jamova 39, SI-1000 Ljubljana, Slovenia}%

\maketitle
\onecolumngrid

\section{Discrete model}
\begin{figure}[htb!]
    \centering
    \includegraphics{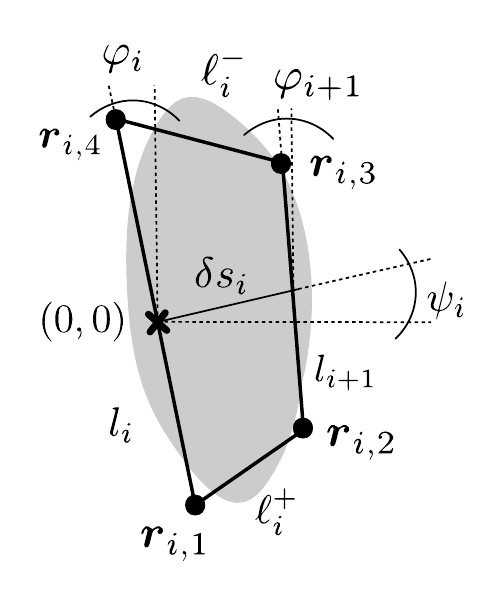}
    \caption{Parametrization of cell shape.}
    \label{fig:cell}
\end{figure}

The shape of $i$th cell is parametrized by the lengths of the two lateral sides $l_i$ and $l_{i+1}$ and by their orientation relative to the vertical ($y$) axis given by angles $\varphi_i$ and $\varphi_{i+1}$ as well as by the cell-midline length (i.e., length of a line segment between midpoints of the lateral sides) denoted by $\delta s_i$. The orientation of the cell is specified by the angle between the midline and the horizontal ($x$) axis denoted by $\psi_i$. In terms of these variables, the positions of the vertices of $i$th cell read (Fig. \ref{fig:cell})
\begin{equation}
	\begin{split}
    \boldsymbol r_{i,1}&=-\frac{l_i}{2}\left (-\sin\varphi_i,\cos\varphi_i,0\right )\>,\\
    \boldsymbol r_{i,2}&=\frac{l_i}{2}\left (-\sin\varphi_i,\cos\varphi_i,0\right )\>,\\
    \boldsymbol r_{i,3}&=\left (\delta s_i\cos\psi_i+\frac{l_{i+1}}{2}\sin\varphi_{i+1},\delta s_i\sin\psi_i-\frac{l_{i+1}}{2}\cos\varphi_{i+1},0\right )\>, \qquad {\rm and}\\
    \boldsymbol r_{i,4}&=\left (\delta s_i\cos\psi_i-\frac{l_{i+1}}{2}\sin\varphi_{i+1},\delta s_i\sin\psi_i+\frac{l_{i+1}}{2}\cos\varphi_{i+1},0\right )\>.
	\end{split}
\end{equation}
In addition, the dimensionless cell area equals unity:
\begin{equation}
    A_i=\frac{1}{2}\sum_{\nu}\left (\boldsymbol r_{i,\nu}\times\boldsymbol r_{i,\nu+1}\right )\cdot\boldsymbol e_z=1\>,
\end{equation}
which allows us to eliminate $\delta s_i$ by expressing it with the remaining variables as 
\begin{equation}
    \delta s_i=\frac{2}{l_i\cos{\left (\psi_i-\varphi_i\right )}+l_{i+1}\cos{\left (\psi_i-\varphi_{i+1}\right )}}\>.
    \label{eq:Ss}
\end{equation}
In dimensionless form, where the units of energy and length are given by $\Gamma_l\sqrt{A_0}$ and $\sqrt{A_0}$, respectively, the energy of a single cell reads
\begin{equation}
    w_i=\frac{\Gamma_b}{\Gamma_l}\ell_i^{+}+\frac{\Gamma_a}{\Gamma_l}\ell_i^{-}+\frac{1}{2}\left (l_i+l_{i+1}\right )\>.
    \label{eq:energy}
\end{equation}
Here
\begin{equation}
    \ell_i^{\pm}=\frac{1}{2}\sqrt{l_i^2+l_{i+1}^2-2l_il_{i+1}\cos{\left(\varphi_{i+1}-\varphi_i\right )}+4s_i\left (s_i\pm d_i\right )}\>,
\end{equation} 
are the lengths of the basal and the apical sides, calculated directly from the vertex positions; here $d_i=l_i\sin{\left (\psi_i-\varphi_i\right )}-l_{i+1}\sin{\left (\psi_i-\varphi_{i+1}\right )}$.

The energy of the whole tissue is a sum of the energies of individual cells  
\begin{equation}
    W=\sum_i^{N} w_i\>.
\end{equation} 
For the reference state, we take the flat configuration where $l_i=\sigma_0^{-1}=\Gamma^{1/2}$, $\varphi_i=0$, and $\psi_i=0$ for all $i$.
\section{Effective elastic moduli}
\subsection{Stretching modulus}
In a flat epithelium (i.e., for $\Delta=0$), the energy per cell reads
\begin{equation}
    w_{\rm flat}=\Gamma \sigma^*+\frac{1}{\sigma^*}\>,
\end{equation}
where $\sigma^*$ is cell width. The energy-minimizing cell width is determined by the balance of forces $\partial w_{\rm flat}/\partial \sigma^*=0$ and reads $\sigma_0=\Gamma^{-1/2}$.

To derive the stretching modulus, we consider a deformed tissue where $\sigma^*=\sigma_0+\delta \sigma$. We Taylor-expand the total energy for $\delta \sigma^*/\sigma_0\ll 1$ to second order so as to obtain
\begin{equation}
    w_{\rm flat}+\delta w_{\rm flat}=\Gamma\left (\sigma_0+\delta \sigma\right )+\frac{1}{\sigma_0+\delta \sigma}
    \approx 2\Gamma^{1/2}+\frac{2\Gamma^{1/2}}{2}\left (\frac{\delta \sigma}{\sigma_0}\right )^2\>.
\end{equation}
By comparing the deformation energy $\delta w_{\rm flat}$ with the elastic energy
\begin{equation}
    \delta w_{\rm flat}=\frac{2\Gamma^{1/2}}{2}\left (\frac{\delta \sigma}{\sigma_0}\right )^2=\frac{K_s}{2}\int_0^{\sigma_0}{\rm d}\sigma\left (\frac{\delta \sigma}{\sigma_0}\right )^2\>,
\end{equation}
we obtain the stretching modulus, which reads $K_s=2\Gamma$. Due to the fixed-cell-area constraint, the equilibrium cell height $l_0=\sigma_0^{-1}=\Gamma^{1/2}$ and thus 
\begin{equation}
    K_s=2l_0^2\>.
\end{equation}
\subsection{Bending modulus}
%

The energy per cell of an epithelium with a constant curvature c and $\Delta=0$ reads
\begin{equation}
    \label{energyTrap}
    w_{\rm trap}=\frac{\Gamma}{l}+l\sqrt{1+\left (\frac{c}{2l}\right )^2}\>,
\end{equation}
From the force balance equations $\partial w/\partial l=0$ and $\partial w/\partial c=0$ we obtain the equilibrium cell height and curvature
\begin{equation}
        l_0=\Gamma^{1/2}
\end{equation}
and 
\begin{equation}
        c_0=0\>,
\end{equation}
respectively. 

Next, we Taylor-expand Eq.~(\ref{energyTrap}) around $c=c_0=0$ and divide it by $\sigma_0=l_0^{-1}$ so as to obtain the energy density $w_{\rm trap}/\sigma_0$. Finally, we compare $w_{\rm trap}/\sigma_0$ with
\begin{equation}
    \frac{w_{\rm trap}}{\sigma_0}=\frac{w_{\rm flat}}{\sigma_0}+\frac{B}{2}c^2\>
\end{equation}
and we find that the bending modulus
\begin{equation}
    B=\frac{1}{4}\>.
    \label{eq:bendingmodulusflat}
\end{equation}
\section{Continuum theory}
Cell-midline length $\delta s_i$ and variables $l_i$, $\varphi_i$, and $\psi_i$ are next written as continuous functions of the distance $\sigma$ along the midline of the reference state: 
\begin{equation}
	\begin{split}
	    \delta s_i&\to \delta s(\sigma)\>,\\
		l_i&\to l(\sigma)\>,\\
		\varphi_i&\to\varphi(\sigma)\>,\qquad {\rm and}\\
		\psi_i&\to\psi(\sigma)\>.
	\end{split}
\end{equation}
In turn, the variables $l_{i+1}$ and $\varphi_{i+1}$ are related to $l(\sigma)$ and $\varphi(\sigma)$ through derivatives $\dot l=\partial_\sigma l(\sigma)$ and $\dot \varphi=\partial_\sigma \varphi(\sigma)$ as 
\begin{equation}
	\begin{split}
		l_{i+1}&\approx l(\sigma)+\sigma_0\dot l(\sigma)\>\qquad {\rm and}\\
		\varphi_{i+1}&\approx\varphi(\sigma)+\sigma_0\dot\varphi(\sigma)\>.
	\end{split}
\end{equation}
In the continuum limit, the sum of cell energies is replaced by integral over the midline length of the reference state $\left (\sum_{i=1}^N\to\int_0^{N\sigma_0}{\rm d}\sigma/\sigma_0\right )$ so that the energy of the tissue is given by 
\begin{equation}
    W=\int_0^{N\sigma_0}\mathcal L\>{\rm d}\sigma=\Gamma^{1/2}\int_0^{N\sigma_0}\left [w\left (l,\varphi,\psi,\dot l,\dot\varphi\right )+\mu\delta s\left (l,\varphi,\psi,\dot l,\dot\varphi\right )\cos\psi \right]{\rm d}\sigma\>.
\end{equation}
The term $\mu\delta s\cos\psi$ is included in the integrand so as to control the end-to-end distance of the tissue, $D=\Gamma^{1/2}\int_0^{N\sigma_0}\delta s\cos\psi\>{\rm d}\sigma$, the Lagrange multiplier $\mu$ being the force along the horizontal ($x$) direction.

We expand the Lagrangian density $\mathcal L$ for small deviations from the reference flat configuration to the second order and obtain
\begin{equation}
	\begin{split}
		\mathcal L\approx &2\Gamma - {\Delta \sqrt{\Gamma} \over 2} \dot\varphi +\delta l^2+{\Gamma \over 2} \left(\psi-\varphi \right)^2 
		+  {2 + \Gamma^2 \over 8 \Gamma} \delta\dot l^2 - {\Delta \over 2 \sqrt{\Gamma}} \delta\dot l\dot\varphi +\\ &+ {1 \over 4}\dot\varphi^2 + \delta l \left({\delta\dot l \over \sqrt{\Gamma}} - {\Delta \over 2 } \dot\varphi\right) +\left(\psi-\varphi \right) \left({\Delta \over 2} \delta\dot l - {\sqrt{\Gamma} \over 2} \dot\varphi \right)+ \\ &+ \mu  \left( 1+ \frac{\psi ^2-(\varphi -\psi )^2}{2}+\frac{\delta \dot{l}^2}{4 \Gamma ^{2}}+\frac{4 \delta l^2+\dot{\varphi} ^2}{4 \Gamma}-\frac{2 \delta l+ \psi \dot{\varphi}}{2 \sqrt{\Gamma} }\right)\>, 
	\end{split}
    \label{eq:lagrangian}
\end{equation}
where $\delta l=l-\sigma_0^{-1}$ is the deviation of the lateral-side length from the reference cell height $\sigma_0^{-1}$. 

The corresponding Euler--Lagrange equations read
\begin{align}
        4 \Gamma ^{3/2} \mu +\left(\Gamma ^3+2 \Gamma +2 \mu \right) \delta \ddot l+2 \Gamma ^2 \Delta  \dot{\psi}-8 \Gamma  (\Gamma +\mu ) \delta l-2 \Gamma ^{3/2} \Delta  \ddot{\varphi}&=0\>, \label{eq:ELE1}\\
        \sqrt{\Gamma } \left\{\Delta  \delta \ddot l+(\Gamma +\mu ) \left[2 \sqrt{\Gamma } (\varphi -\psi)+\dot{\psi}\right]\right\}-(\Gamma +\mu ) \ddot{\varphi}&=0\>,\qquad {\rm and}
         \label{eq:ELE2}\\
        \sqrt{\Gamma } \left[\Delta  \delta \dot l-2 (\Gamma +\mu ) \varphi+2 \Gamma  \psi \right]-(\Gamma +\mu ) \dot{\varphi}&=0\>.
        \label{eq:ELE3}
\end{align}
%
%
%
\section{Wrinkling}
We write the 
variables $\delta l(\sigma), \varphi(\sigma)$ and $\psi(\sigma)$ as sums of Fourier modes characterized by the wave number $q=2\pi/\lambda$: 
\begin{equation}
	\begin{split}
	\delta l(\sigma)&= \frac{1}{2\pi}\int \delta \tilde l (q)\exp({\rm i}q\sigma){\rm d}q\>, \\
    \varphi(\sigma)&=\frac{1}{2\pi}\int \tilde\varphi (q)\exp({\rm i}q\sigma){\rm d}q\>, \qquad{\rm and }\\
    \psi(\sigma)&=\frac{1}{2\pi}\int \tilde\psi (q)\exp({\rm i}q\sigma){\rm d}q\>.
    \end{split}
\end{equation}

Assuming that variables $ \varphi(\sigma)$ and $\psi(\sigma)$ have a dominant mode at $q=q_0$ with amplitudes $\tilde\varphi (q_0)$ and $\tilde\psi (q_0)$, respectively, whereas $\delta l(\sigma)$ has dominant modes at both $q=q_0$ and $q=0$ with amplitudes $\delta \tilde l(q_0)$ and $\delta \tilde l(0)$, respectively, we obtain a system of three equations

%
\begin{align}
    \begin{split}
        \exp(-{\rm i} q_0 \sigma) \left[8 \Gamma  \delta \tilde l(0) (\Gamma +\mu )-4 \Gamma ^{3/2} \mu\right]+\\+8 \Gamma  \delta \tilde l(q_0) (\Gamma +\mu )+\delta \tilde l(q_0) q_0^2 \left(\Gamma ^3+2 \Gamma +2 \mu \right)-2 \Gamma ^{3/2} \Delta  q_0 \left[q_0 \tilde\varphi (q_0)+{\rm i} \sqrt{\Gamma } \tilde\psi (q_0)\right]&=0 
        \label{eq:ELE2Z}
    \end{split}    
        \\
        \sqrt{\Gamma } \Delta  \delta \tilde l(q_0) q_0^2-(\Gamma +\mu ) \left[2 \Gamma  (\tilde\varphi (q_0)-\tilde\psi (q_0))+q^2 \tilde\varphi (q_0)+{\rm i} \sqrt{\Gamma } q_0 \tilde\psi (q_0)\right]&=0\>\qquad{\rm and}\label{eq:ELE2Y}\\
        {\rm i} q_0 \left[\sqrt{\Gamma } \Delta  \delta \tilde l(q_0)-\tilde\varphi (q_0) (\Gamma +\mu )\right]-2 \sqrt{\Gamma } \left[\tilde\varphi (q_0) (\Gamma +\mu )-\Gamma  \tilde\psi (q_0)\right]=0\>.
        \label{eq:ELE2X}
\end{align}
From Eq.~(\ref{eq:ELE2Z}) we find that
\begin{equation}
   \delta \tilde l(0)=\frac{\sqrt{\Gamma } \mu }{2 (\Gamma +\mu )},
   \label{eq:l_constant}
\end{equation}
and we express $\delta \tilde l(q_0)$ 
\begin{equation}
    \delta \tilde l(q_0)=\frac{2 \Gamma ^{3/2} \Delta  q_0 \left[q_0 \tilde\varphi (q_0) +{\rm i} \sqrt{\Gamma } \tilde\psi (q_0)\right]}{8 \Gamma  (\Gamma +\mu )+q_0^2 \left(\Gamma ^3+2 \Gamma +2 \mu \right)}
\end{equation}
and insert it into Eq.~(\ref{eq:ELE2Y}) and (\ref{eq:ELE2X}). From Eq.~(\ref{eq:ELE2Y}) we then express $\tilde\varphi (q_0)$ as 
\begin{equation}
    \tilde \varphi (q_0)=\frac{-2 \Gamma  \tilde\psi (q_0)(\Gamma +\mu )+i \sqrt{\Gamma } q_0 \tilde\psi (q_0) (\Gamma +\mu )-2 {\rm i} \Gamma ^{5/2} \Delta ^2 q_0^3 \tilde\psi (q_0)/\left[8 \Gamma  (\Gamma +\mu )+q_0^2 \left(\Gamma ^3+2 \Gamma +2 \mu \right)\right]}{-2 \Gamma  (\Gamma +\mu )-q_0^2 (\Gamma +\mu )+2 \Gamma ^2 \Delta ^2 q_0^4/\left[8 \Gamma  (\Gamma +\mu )+q_0^2 \left(\Gamma ^3+2 \Gamma +2 \mu \right)\right]}
\end{equation}
and insert it into Eq.~(\ref{eq:ELE2X}). This yields an expression $\tilde\psi(q_0)f(q_0)=0$, where $f(q_0)$ is en expression containing only variables $q_0$, $\Gamma$, $\Delta$, and $\mu$. If $\tilde\psi(q_0)\neq0$, this allows us to derive an algebraic equation for the wave number $q_0$ which reads
%
%
\begin{equation}
    \label{poly}
    q_0^4 +q_0^2 \frac{4 \Gamma   (\Gamma +\mu ) \left[\Gamma ^2 \left(\Delta ^2-2\right)+\left(\Gamma ^2+2\right) \Gamma  \mu +4 \mu ^2\right]}{(\mu -\Gamma ) \left[\Gamma ^2 \left(\Gamma ^2-2 \Delta ^2+2\right)+\left(\Gamma ^2+4\right) \Gamma  \mu +2 \mu ^2\right]}+\frac{32 \Gamma ^2 \mu  (\Gamma +\mu )^2}{(\mu -\Gamma ) \left[\Gamma ^2 \left(\Gamma ^2-2 \Delta ^2+2\right)+\left(\Gamma ^2+4\right) \Gamma  \mu +2 \mu ^2\right]}=0\>.
\end{equation}
By comparing this equation to Eq.~(7) from Ref.~\cite{brau13}, we find that the modulus of the $q_0^2$ term is the effective in-plane force $\widetilde F$ and that third term on the left-hand side is the effective stiffness of a phantom substrate $\widetilde K$. Both $\widetilde F$ and $\widetilde K$ are functions of $\Delta$, $\Gamma$, and $\mu$.

\subsection{Lagrange multiplier}
Equation~(\ref{poly}) has a double real solution at the critical value of the Lagrange multiplier, 
\begin{equation}
    \mu=\mu_c(\Gamma,\Delta)\>,
\end{equation}
which is a solution of a 4th-order polynomial equation:
\begin{multline}
        16 \Gamma ^8 \Delta ^4-64 \Gamma ^8 \Delta ^2+64 \Gamma ^8+\mu  \left(32 \Gamma ^9 \Delta ^2+64 \Gamma ^9+32 \Gamma ^7 \Delta ^4-320 \Gamma ^7 \Delta ^2+256 \Gamma ^7\right)+ \\ +\mu ^2 \left(16 \Gamma ^{10}+64 \Gamma ^8 \Delta ^2+192 \Gamma ^8+16 \Gamma ^6 \Delta ^4-64 \Gamma ^6 \Delta ^2+384 \Gamma ^6\right)+ \\ +\mu ^3 \left(32 \Gamma ^9+32 \Gamma ^7 \Delta ^2+192 \Gamma ^7+576 \Gamma ^5 \Delta ^2+256 \Gamma ^5\right)+\mu ^4 \left(16 \Gamma ^8+64 \Gamma ^6+384 \Gamma ^4 \Delta ^2+64 \Gamma ^4\right)=0\>.
\end{multline}

%
\subsection{Generalized in-plane force}
At the onset of the elastic instability (i.e., at $\mu=\mu_c$),
the generalized in-plane force  reads
\begin{equation}
    \widetilde F_c (\Gamma, \Delta) = \left\{
        \begin{array}{ll}
            \displaystyle{\frac{4\Gamma\left (2-\Delta^2\right )}{2-2\Delta^2+\Gamma^2}}, & \quad \left |\Delta\right |<\Delta_c \\
            \displaystyle{\frac{4 \Gamma  (\Gamma +\mu_c ) \left(\Gamma ^2 \left(\Delta ^2-2\right)+\left(\Gamma ^2+2\right) \Gamma  \mu_c +4 \mu_c ^2\right)}{(\mu_c -\Gamma ) \left[\Gamma ^2 \left(\Gamma ^2-2 \Delta ^2+2\right)+\left(\Gamma ^2+4\right) \Gamma  \mu_c +2 \mu_c ^2\right]}}, & \quad \left |\Delta\right |\geq\Delta_c
        \end{array}
    \right.\>.
\end{equation}
\subsection{Effective stiffness of a phantom substrate}
At the onset of the elastic instability (i.e., at $\mu=\mu_c$), the effective stiffness of the phantom substrate reads
\begin{equation}
    \widetilde K_c (\Gamma, \Delta)= \left\{
        \begin{array}{ll}
            0, & \quad \left |\Delta\right |<\Delta_c \\
            \displaystyle{\frac{32 \Gamma ^2 \mu_c  (\Gamma +\mu_c )^2}{(\mu_c -\Gamma ) \left[\Gamma ^2 \left(\Gamma ^2-2 \Delta ^2+2\right)+\left(\Gamma ^2+4\right) \Gamma  \mu_c +2 \mu_c ^2\right]}}, & \quad \left |\Delta\right |\geq\Delta_c
        \end{array}
    \right.\>.
\end{equation}
%
\subsection{Wavelength}
The double real solution of Eq.~(\ref{poly}), which appears at $\mu=\mu_c$, yields the wavelength of wrinkles 
\begin{equation}
		\lambda_0 = {2\pi \over q_0}=\sqrt{2} \pi \sqrt{\frac{(\Gamma -\mu_c ) \left[\Gamma ^2 \left(\Gamma ^2-2 \Delta ^2+2\right)+\left(\Gamma ^2+4\right) \Gamma  \mu_c +2 \mu_c ^2\right]}{\Gamma  (\Gamma +\mu_c ) \left[\Gamma ^2 \left(\Delta ^2-2\right)+\left(\Gamma ^2+2\right) \Gamma  \mu_c +4 \mu_c ^2\right]}}\>.
    \label{q_analitical}
\end{equation}
%
\subsection{Critical behavior}
Close to the critical point $\left |\Delta\right |=\sqrt{2}$, the polynomial equation for $\mu_c$ can be recast as a linear equation by neglecting higher-order terms. This equation yields
\begin{equation}
    \mu_c (\Gamma, \Delta) = -\frac{\Gamma  \left(\Delta ^2-2\right)^2}{2 \left(\Gamma ^2 \left(\Delta ^2+2\right)+\Delta ^4-10 \Delta ^2+8\right)}\>.
\end{equation}
Just above the critical point (i.e., for $\left |\Delta\right |\gtrsim\Delta_c$), this result simplifies to 
\begin{equation}
    \mu_c(\Gamma, \Delta)\approx-\frac{\Gamma^{1/2}}{\Gamma^2-2}\left (\left |\Delta\right |-\Delta_c\right )^2\>.
\end{equation}
Around the critical point $\left |\Delta\right |=\sqrt{2}$, the critical generalized in-plane force reads 
\begin{equation}
    \widetilde F_c(\Gamma, \Delta)\approx\frac{8\sqrt{2}\Gamma}{2-\Gamma^2}\left (\left |\Delta\right |-\Delta_c\right )\>.
\end{equation}
Just above the critical point, the effective stiffness of a phantom substrate
\begin{equation}
    \widetilde K_c(\Gamma, \Delta)\approx\frac{32\Gamma^2}{\left (\Gamma^2-2\right )^2}\left (\left |\Delta\right |-\Delta_c\right )^2\>,
\end{equation}
whereas the wavelength
%
\begin{equation}
    \lambda_0 = {\pi \sqrt{\Gamma - 2/\Gamma} \over 2^{1/4}} \left(\left |\Delta\right | - \Delta_c\right)^{-1/2}.
\end{equation}

\section{Vertex model}
\label{sec:vertexmodel}
\subsection{Implementation}
In the vertex model, the tissue is represented by a chain of quadrilaterals representing cell cross sections. The basal, apical, and lateral cell membranes are represented by edges carrying line tensions $\Gamma_b$, $\Gamma_a$, and $\Gamma_l$, respectively. Cell shapes are parametrized by positions of the four vertices $\boldsymbol r_{i,1}=(x_{i,1},y_{i,1})$, $\boldsymbol r_{i,2}=(x_{i,2},y_{i,2})$, etc. 

The dimensionless energy of the system reads
\begin{equation}
	W=\sum_i \left[\kappa_A\left (A_i-1\right )^2+\sum_{\nu} \gamma_{i, \nu} l_{i, \nu}\right]\>,
	\label{eq:energyXXX}
\end{equation}
where the sum goes over all cells $i$. The dimensionless line tension $\gamma_i$ equals $\beta$ on the basal side, $\alpha$ on the apical side, and $1/2$ on the lateral sides;
here $\beta=\Gamma_b/\Gamma_l$ and $\alpha=\Gamma_a/\Gamma_l$.
%
%
In Eq.~(\ref{eq:energyXXX}), $\kappa_A$ is the compressibility modulus of the cells; in this study, $\kappa_A=100$ which describes nearly incompressible cells. 

The energy of the tissue is minimized using the gradient descent method. In particular, conservative forces acting on vertices are given by $\boldsymbol F_{i,\nu}=-\nabla_{i,\nu} W$, 
where $\nabla_{i,\nu}={\rm d}/{\rm d}\boldsymbol r_{i,\nu}$
and overdamped dynamics are considered for vertex motion:
\begin{equation}
    \label{gradDesc}
	\frac{{\rm d}\boldsymbol{r}_{i,\nu}}{{\rm d}t}=\boldsymbol F_{i,\nu}=-\sum_j\left[2\kappa_A\left (A_j-1\right )\nabla_{i,\nu}A_j+\sum_{\mu}\gamma_{j, \mu}\nabla_{i,\nu}l_{j, \mu}\right]\>.
\end{equation}
(The damping coefficient has been absorbed in the suitably redefined time.) Edge lengths and cell areas are given by $l_{j,\mu}=\left |\boldsymbol r_{j,\mu+1}-\boldsymbol r_{j,\mu}\right |$ and $A_j=(1/2)\sum_{\mu}\left (\boldsymbol r_{j,\mu}\times\boldsymbol r_{j,\mu+1}\right )\cdot\boldsymbol e_z$, respectively, where $\boldsymbol e_z=(0,0,1)$.
\subsection{Simulations}
\label{sub_initial_c}
For every pair of $\Gamma$ and $\Delta$, the system is integrated for 95 different cell numbers $N={5, 6, 7 ... 100}$, assuming periodic boundary conditions. The initially flat cell configuration of length $N\sigma_0$ is compressed by $\epsilon=\delta \sigma/\sigma_0$, which is varied between $0.01$ and $-0.1$ in steps of 0.005
so as to find the size of the simulation box corresponding to the critical in-plane force at which the instability occurs. The initial configuration is periodically perturbed so as to ensure a more efficient convergence of the system towards the energy minimum. The system of differential equations Eq.~(\ref{gradDesc}) is solved using an explicit Euler scheme with a time step of $\Delta t=10^{-3}$. 

\begin{figure}[h]
    \centering
    \includegraphics[width=7cm]{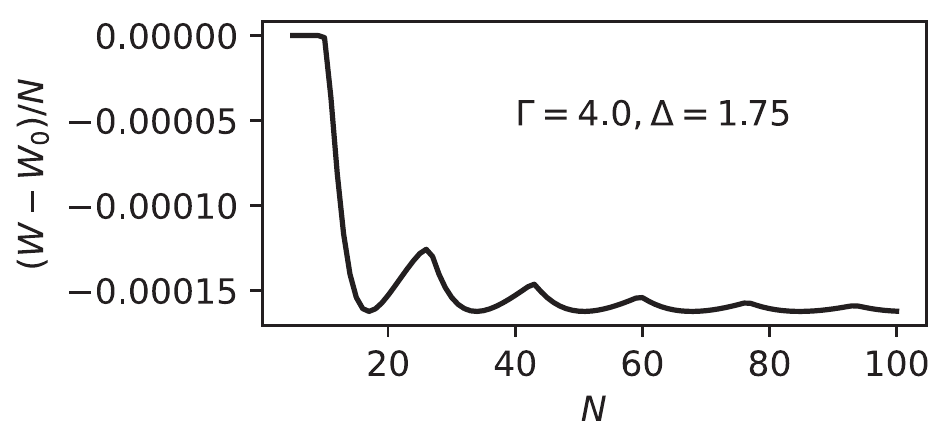}
    \caption{Example of energy per cell as a function of cell number in a tissue with $\Gamma=4$ and $\Delta=1.75$; $W_0$ is the energy of a flat model tissue at these $\Gamma$ and $\Delta$. Strain $\epsilon=0.01.$ 
    }
    \label{fig:WN}
\end{figure}

Among the final shapes at the critical strain $\epsilon$ we identify the one with the lowest energy per cell, its wavelength being $\lambda_0(\Gamma, \Delta)=N_0\sigma_0(1-\epsilon)$ where $N_0(\Gamma, \Delta,\epsilon)$ is the equilibrium number of cells in one waveform. The energy-degenerate states shown in Fig.~\ref{fig:WN} correspond to multiples of the smallest equilibrium cell number $N_0$.

%
%

\subsection{Equilibrium shapes at large compressive strain}
As thin sheets on elastic substrate undergo period-doubling bifurcation, simulate optimal tissue shapes at large relative compression (up to $\epsilon=0.4$). 
We investigate the presence of period-doubling bifurcation by considering tissues comprised of $N$ cells, where $N$ is varied between $N_0(\Gamma, \Delta, \epsilon)$ and $2.2N_0(\Gamma, \Delta,\epsilon)]$; $N_0(\Gamma, \Delta, \epsilon)$ is the optimal number of cells in a single-mode wrinkled state at $\Gamma$, $\Delta$, and strain $\epsilon$. In case of a period doubling, these tissues would relax into a configuration with one period containing two modes (i.e., two $q$s). Instead, we find that they relax into a configuration with two periods of a single (original) mode.

\begin{figure}[h]
    \centering
    \includegraphics[width=6cm]{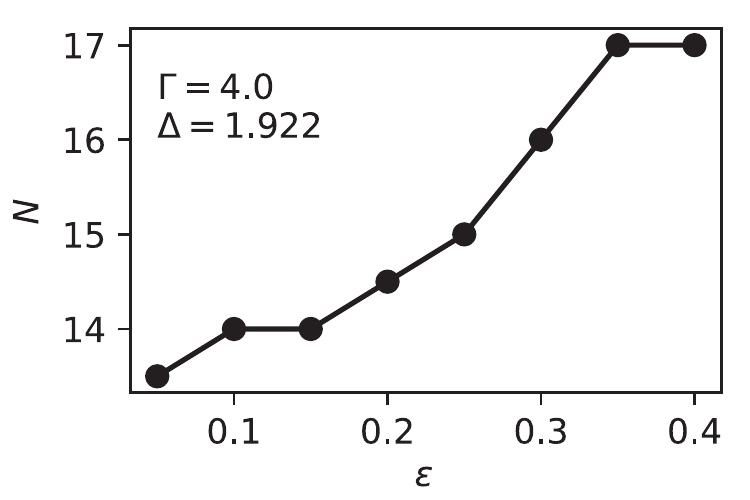}
    \caption{Number of cells $N$ in one waveform 
    as a function of compressive strain $\epsilon$ at $\Gamma=4$ and  $\Delta=1.922$. }
    \label{fig:N_compression}
\end{figure}

The results show that the equilibrium cell number per waveform $N_0$ grows with the strain, but the dependence is rather weak as illustrated in Fig.~\ref{fig:N_compression}.


%

\begin{figure}[h]
    \centering
    \includegraphics[width=8cm]{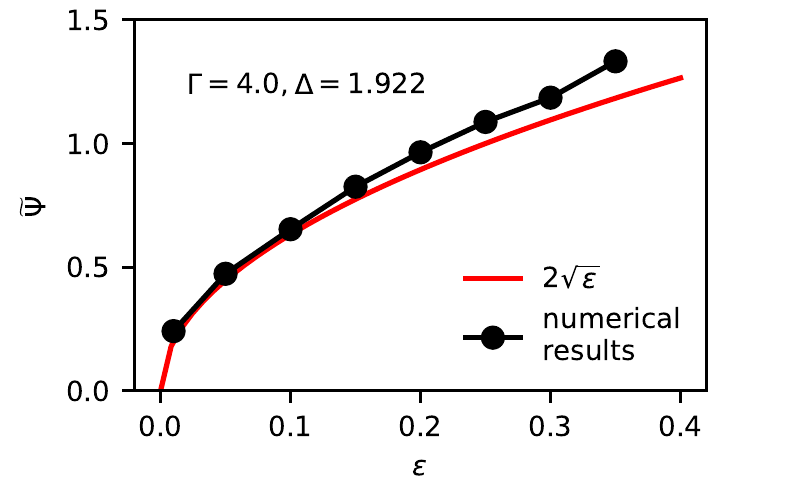}
    \caption{Dependence of amplitude of the deformation $\widetilde\psi$ on the strain $\epsilon$ for $\Gamma=4$ and $\Delta=1.922$.}
    \label{fig:highcompression}
\end{figure}

The amplitude of wrinkles $\psi_0$ grows with the strain $\epsilon$. For a thin sheet with contour length $L_0$, compressed by a relative compression $\epsilon$, the incompressibility constraint reads
\begin{equation}
    L_0\left(1-\epsilon\right)=\int_{0}^{L_0}\cos\psi(s)\mathrm{d}s \approx L_0 - {1\over 2}\int_0^{L_0} \psi(s)^2\mathrm{d}s.
    \label{eq:incompressibility}
\end{equation}
Assuming periodic solution $\psi(\sigma)=\widetilde\psi\sin\left({2\pi \sigma/ \lambda_0}\right)$, where $\lambda_0=L_0\left(1-\epsilon\right)$ and solving Eq.~(\ref{eq:incompressibility}) up to the leading order in $\epsilon$ gives the relation $\widetilde\psi=2\sqrt{\epsilon}$. Our numerical results show a slight deviation from this theoretical prediction, since the tissue-midline length is not fixed, but incompressibility is assumed for areas of cell cross sections (Fig.~\ref{fig:highcompression}).
\begin{figure}[h!]
    \centering
    \includegraphics{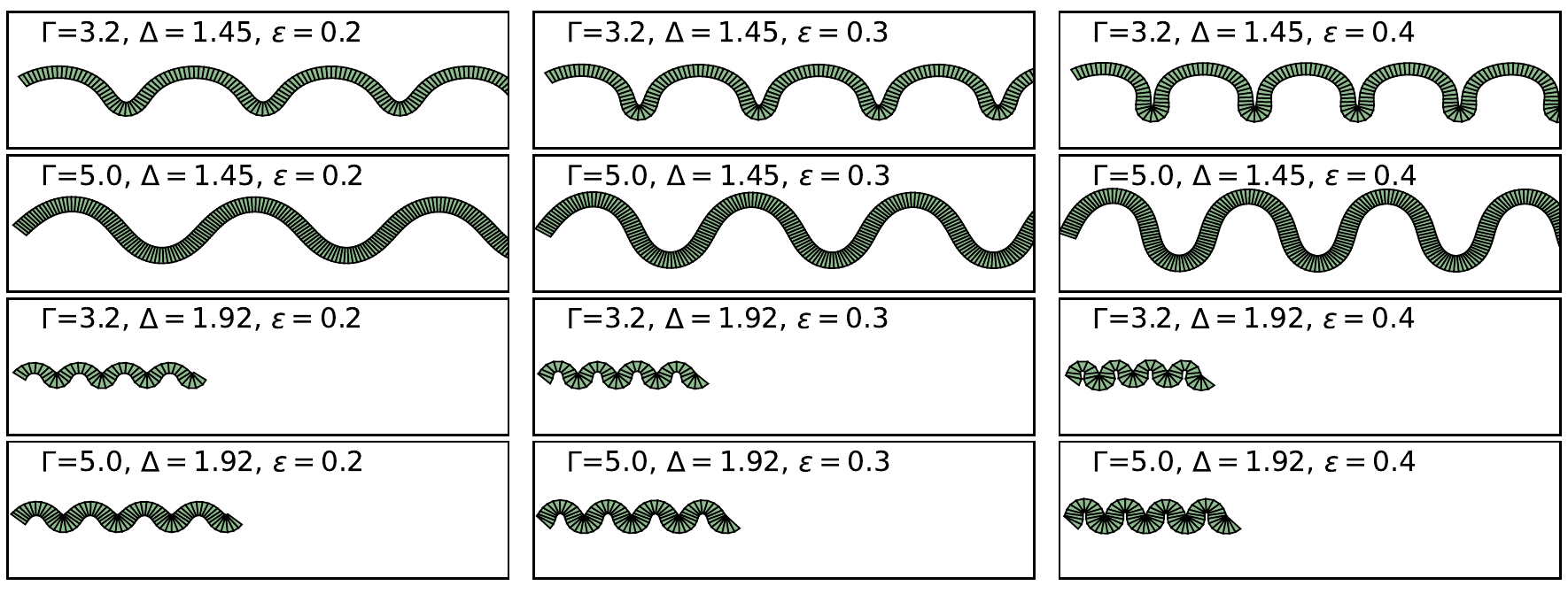}
    \caption{Relaxed shapes of tissues under high relative compression $\epsilon$ for some extreme values of $\Gamma, \Delta$ composed of $5N_0$ cells, where $N_0$ is the optimal number of cells in one waveform at small compression $\epsilon=0.01$. Even at compression of $\epsilon=0.4$ the tissues remain in wrinkle state and do not undergo wrinkle-to-fold transition in contrast to thin sheet on liquid substrate.}
    \label{fig:high_compression}
\end{figure}
%
%



{\bf No wrinkle-to-fold transition.} We simulated tissues composed of $5N_0$ cells, where $N_0$ is the optimal number of cells in the wrinkled configuration and found no folded states at high compression~(Fig.~\ref{fig:high_compression}). This indicates that even at relative compression $\epsilon=0.4$ epithelial tissues do not undergo wrinkle-to-fold transition in contrast to thin elastic plates attached to liquid substrates.
\section{Two-cell model}

We model the waveform by two oppositely curved segments: (1)~the groove that contains $N_1$ identical cells of height $h_1$ and (2)~the crest that contains $N_2=N-N_1$ identical cells of height $h_2$. Further, we assume that the tissue midline assumes the shape of two arcs with opposite curvatures $c_1=-1/R$ and $c_2=1/R$ but with identical angles both denoted by $\phi$.

The total energy of the waveform reads
\begin{equation}
    W=N_1w_1+(N-N_1)w_2\>,
\end{equation}
where
\begin{equation}
    w_1\approx\frac{\Gamma}{h_1}-\frac{\Delta}{2R}+h_1+\frac{1}{8R^2h_1}
\end{equation}
and
\begin{equation}
    w_2\approx\frac{\Gamma}{h_2}+\frac{\Delta}{2R}+h_2+\frac{1}{8R^2h_2}
\end{equation}
are the energies of the groove cell and the crest cell, respectively.

The cells from the two cell populations are coupled by geometric constraints: $\phi R=N_1s_1=N_2s_2$, where $s_1=1/h_1$ and $s_2=1/h_2$ are widths of groove cell and crest cell, respectively. This constraint yields
\begin{equation}
    N_1=N\frac{h_1}{h_1+h_2}
\end{equation}
and the total energy per cell reads
\begin{equation}
    \frac{W}{N}=\frac{h_1}{h_1+h_2}w_1+\left (1-\frac{h_1}{h_1+h_2}\right )w_2=\frac{1+2R[2R(h_1^2+h_2^2)+\Delta(h_2-h_1)+4R\Gamma]}{4(h_1+h_2)R^2}\>.
\end{equation}
Evaluating the energy at the minimum, where
\begin{equation}
    h_1=\frac{R\Delta+\sqrt{R^2(2-\Delta^2+16R^2)}}{4R^2}
\end{equation}
and
\begin{equation}
    h_2=\frac{-R\Delta+\sqrt{R^2(2-\Delta^2+16R^2)}}{4R^2}\>,
\end{equation}
yields
\begin{equation}
    \frac{W}{N}=\sqrt{\frac{16R^2\Gamma+2-\Delta^2}{4R^2}}\>.
\end{equation}
For $\Delta^2$ close to $2$
\begin{equation}
    \label{eq:enDiff}
    \frac{W}{N}\approx 2\sqrt{\Gamma}+\frac{2-\Delta^2}{16R^2\sqrt{\Gamma}}\>.
\end{equation}
Since $2\sqrt{\Gamma}$ is the energy per cell of a flat tissue, the waveform 
consisting of two segments with opposite curvatures and different cell heights becomes energetically favorable at $\left |\Delta\right |=\sqrt{2}$, which agrees with the critical point obtained from the elasticity theory and the vertex model. 

Furthermore, the difference of cell heights in the groove and the crest of the model waveform
\begin{equation}
    h_1-h_2=\frac{\Delta}{2R}\>.
\end{equation}
also agrees with the amplitude of the cell-height modulation obtained within the elasticity theory and the vertex model. 
\section{Energy of wrinkled state}
From the numerically obtained waveforms of the wrinkled tissues we conclude that they can be described by the following ans\"atze for $\psi(\sigma)$, $\varphi(\sigma)$, and $\delta l(\sigma)$:
\begin{equation}
	\begin{split}
	\psi(\sigma)&= \widetilde\psi \sin(q\sigma)\>,\\
	\varphi(\sigma)&=\widetilde\varphi(\sigma) \sin(q\sigma)\> \qquad {\rm and}\\
	\delta l(\sigma)&= \frac{\sqrt{\Gamma } \mu }{2 (\Gamma +\mu )}+\delta \widetilde{l}(\sigma) \cos(q\sigma)\>,
	\end{split}
\end{equation}
where $\widetilde\psi$ is a number whereas $\widetilde\varphi(\sigma)$ and $\widetilde{l}(\sigma)$ are yet unknown functions of $\sigma$ and the constant term in expression $\delta l(\sigma)$ is taken from the solution of Eq. (\ref{eq:l_constant}). We insert these ans\"atze
into the Euler--Lagrange equations [Eqs.~(\ref{eq:ELE1})-(\ref{eq:ELE3})], and
\begin{figure}[h]
    \centering
    \includegraphics{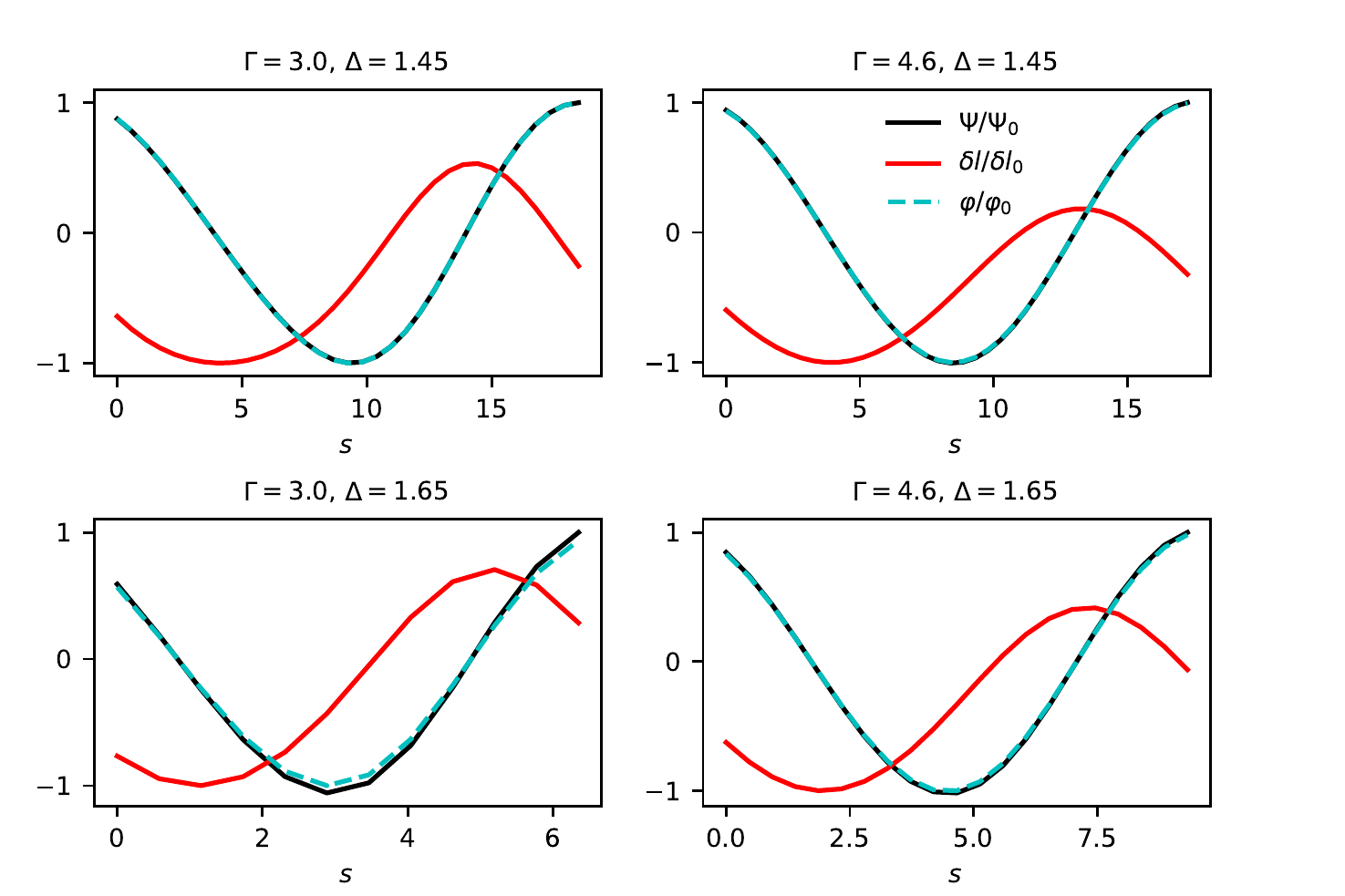}
    \caption{Normalized plots of numerically calculated profiles of $\psi(\sigma), \delta l(\sigma)$ and $\varphi(\sigma)$ for four tissues with different values $\Gamma$ and $\Delta$ (indicated in each panel). Each profile of $\psi(\sigma), \delta l(\sigma)$ and $\varphi(\sigma)$ is normalized by its maximal absolute value, denoted by $\psi_0=\widetilde\psi, \delta l_0$ and $\varphi_0$, respectively. Plots indicate that in a tissue with periodic solutions for $\psi(\sigma), \delta l(\sigma)$ and $\varphi(\sigma)$, the phase shift between $\psi(\sigma)$ and $\varphi(\sigma)$ is approximately 0, whereas the phase shift between $\psi(\sigma)$ and $\delta l(\sigma)$ is approximately $\pi/2$.}
    \label{fig:my_label2}
\end{figure}
we express the amplitudes $\widetilde\varphi$ and $\delta \widetilde l$ 
in terms of $\widetilde\psi$:
\begin{equation}
    \widetilde\varphi(\sigma)=\widetilde{\psi}\frac{2 \Gamma   (\Gamma +\mu )+\sqrt{\Gamma } q  \cot (q \sigma) \left[\Gamma +\mu -{2 \Gamma ^2 \Delta ^2 q^2 \over 8 \Gamma  (\Gamma +\mu )+q^2 \left(\Gamma ^3+2 \Gamma +2 \mu \right)}\right]}{2 \Gamma  (\Gamma +\mu )+q^2 (\Gamma +\mu )-\frac{2 \Gamma ^2 \Delta ^2 q^4}{8 \Gamma  (\Gamma +\mu )+q^2 \left(\Gamma ^3+2 \Gamma +2 \mu \right)}}
\end{equation}
and
\begin{equation}
    \delta \widetilde l (\sigma)=\frac{2 \Gamma ^{3/2} \Delta  q \left[\sqrt{\Gamma } \widetilde{\psi}+q \widetilde{\varphi}(\sigma) \tan (q s)\right]}{8 \Gamma  (\Gamma +\mu )+q^2 \left(\Gamma ^3+2 \Gamma +2 \mu \right)}.
\end{equation}
After inserting the complete solutions for $\psi(\sigma)$, $\varphi(\sigma)$, and $\delta l(\sigma)$ into the Lagrangian $\mathcal{L}$ 
%
%
and averaging it over one wavelength $\lambda=2\pi/q$ as $L=(1/\lambda)\int_{0}^{\lambda}\mathcal L(\sigma) \mathrm{d}\sigma$, we Taylor-expand the expression to fourth order in $q$ around $q=0$ and to first order in $\Delta$ around $\Delta=\Delta_c$. We define $\delta L = L-2\Gamma$, where $2\Gamma$ is the energy of a flat tissue and obtain

\begin{equation}
    \frac{\delta L}{\widetilde\psi^2}\approx -\frac{\mu}{4}+\frac{\left(\Delta -\sqrt{2}\right) }{8 \sqrt{2} } q^2 +\frac{ \left(\sqrt{2} \left(\Gamma ^2+2\right) \Delta -\Gamma ^2-6\right)}{128 \Gamma }q^4\>.
\end{equation}